# Structure of polarimetric purity of a Mueller matrix and sources of depolarization

**José J. Gil**

*Facultad de Educación. Universidad de Zaragoza. c/ Pedro Cerbuna 12, 50009 Zaragoza, Spain*

ppgil@unizar.es

**Abstract**

The depolarization properties of a medium with associated Mueller matrix **M** are characterized through two complementary sets of parameters, namely 1) the three indices of polarimetric purity (IPP), which are directly linked to the relative weights of the spectral components of **M** and provide complete information on the structure of polarimetric randomness, but are insensitive to the specific polarimetric behaviors that introduce the lack of randomness, and 2) the set of three components of purity (CP), constituted by the polarizance, the diattenuation and the degree of spherical purity. The relations between these sets of physical invariant quantities are studied by means of their representation into a common purity figure. Furthermore, the polarimetric properties of a general Mueller matrix **M** are parameterized in terms of sixteen meaningful quantities, three of them being the IPP, which together with the CP provide complete information on the integral depolarization properties of the medium.

## 1. Introduction

Mueller polarimetry is applied for the analysis and study of a great variety of material samples in a continuously increasing number of areas of science, engineering, industry, medicine, etc. Nevertheless, the interpretation of the measured Mueller matrices, as well as the extraction of physical parameters containing decoupled information on the nature and properties of the sample is not a straightforward task. The polarimetric features of a material medium combine, in a complicated manner, polarizing, diattenuating, retarding and depolarizing effects. Therefore, the optimum knowledge of the structure of Mueller matrices is strongly required for the exploitation of polarimetric measurements.

This work is devoted to the study of the depolarization properties of a material sample and describes how the two alternative approaches called the *indices of polarimetric purity* (hereafter IPP) [1] and the *components of purity* (hereafter CP) [2] are mutually related and can be jointly analyzed by means of their graphic representation into an common *purity figure*. For the sake of self-consistency and readability, this article is organized into the following sections. The present Section is devoted to the introduction of the main general notions involved in the further developments as well as the necessary conventions, terminology and notation. Sections 2 and 3 deal respectively with the definition and interpretation of the IPP and the CP of a medium with a given associated Mueller matrix **M**. There are some polarimetric properties of a given material medium (including both IPP and CP) that remain invariant when the medium is serially combined with retarders, so that all so-called invariant-equivalent Mueller matrices, which constitute the subject of Section 4, share the same location in the purity figure. The purity figure, where the different types of Mueller matrices are represented according to the values of their CP and their IPP is studied in Section 5. Section 6 is dedicated to the analysis of the purity figures for the type-I and type-II *canonical depolarizers* [3]. These kinds of matrices are of special interest because they are representative of some intrinsic depolarizing properties of a given Mueller matrix **M**. Section 7 deals with a parameterization of **M** in terms of meaningful phenomenological parameters that highlights the fact that depolarization is fully characterized, in quantity and quality, by means of five parameters, namely the three IPP, the diattenuation and the polarizance. Finally, Section 8 summarizes and discusses the main results and conclusions.

The interaction of a fully polarized beam with a medium that behaves as linear, deterministic, homogeneous and non-depolarizing, can be represented by means of the transformation $\mathbf{\Phi}' = \mathbf{T}\mathbf{\Phi}\mathbf{T}^{\dagger}$, where $\mathbf{\Phi}$, $\mathbf{\Phi}'$ are the input and output polarization matrices respectively, $\mathbf{T}$ is the Jones matrix that characterizes the polarimetric properties of the nondepolarizing medium for the given interaction conditions, and the dagger indicates conjugate transposed. This basic polarimetric interaction can also be expressed as $\mathbf{s}' = \mathbf{M}_J\,\mathbf{s}$ in terms of the corresponding input and output Stokes vectors ($\mathbf{s}$ and $\mathbf{s}'$ respectively) and the Mueller-Jones matrix (or *pure Mueller matrix*) $\mathbf{M}_J(\mathbf{T})$.
The physical polarimetric quantities that characterize this kind of *pure* systems can be easily identified by means of product (serial) decompositions of $\mathbf{T}$ (or $\mathbf{M}_J$), as for instance the polar decomposition [4] or the general serial decomposition [5,6]. Thus, the diattenuation, the polarizance and the retardance exhibited by the pure medium are easily decoupled and interpreted.

In general, the polarimetric behavior of a medium can be considered as a sort of incoherent convex sum, or ensemble average, of nondepolarizing interactions, in such a manner that a physical Mueller matrix can be expressed as a convex sum of pure Mueller matrices [7,8].

Leaving aside passivity constraints (i.e. the fact that naturally occurring phenomena do not increase the intensity of the incoming electromagnetic beams), two alternative, but equivalent, ways for the general characterization of Mueller matrices have been reported *1)* the nonnegativity of the Hermitian matrix $\mathbf{H}$ (covariance or coherency matrix) associated with a given physical Mueller matrix (Cloude's criterion) [9,10], and *2)* the nonnegativity of the N-matrix $\mathbf{G}\mathbf{M}^T\mathbf{G}\mathbf{M}$, where $\mathbf{G} \equiv \mathrm{diag}(1,-1,-1,-1)$ is the Minkowski metric [11-15].

Thus, the structure of a general (depolarizing) Mueller matrix is rather more complicated than that of a pure Mueller matrix. In fact, unlike the transmittance (or reflectance) for unpolarized input light (hereafter called *mean intensity coefficient*), and unlike the diattenuation and polarizance properties, which are easily identified and defined from the given Mueller matrix $\mathbf{M}$ [16,17], the identification and parameterization of the depolarization properties are not so straightforward and require particular study and analysis.

Let us first recall that any Mueller matrix $\mathbf{M}$ can be expressed as [18]

$$\mathbf{M} = m_{00}\begin{pmatrix} 1 & \mathbf{D}^T \\ \mathbf{P} & \mathbf{m} \end{pmatrix},$$

$$\mathbf{D} \equiv \frac{1}{m_{00}}\left(m_{01}, m_{02}, m_{03}\right)^T, \quad \mathbf{P} \equiv \frac{1}{m_{00}}\left(m_{10}, m_{20}, m_{30}\right)^T, \quad \mathbf{m} \equiv \frac{1}{m_{00}}\begin{pmatrix} m_{11} & m_{12} & m_{13} \\ m_{21} & m_{22} & m_{23} \\ m_{31} & m_{32} & m_{33} \end{pmatrix}, \tag{1}$$

where $\mathbf{D}$ and $\mathbf{P}$ are the respective *diattenuation vector* and *polarizance vector* of $\mathbf{M}$. The absolute values of these vectors are the *diattenuation* $D \equiv |\mathbf{D}|$ and the *polarizance* $P \equiv |\mathbf{P}|$. Both polarizance $P$ and diattenuation $D$ have dual nature depending on the direction of propagation of light (forward or reverse) [10,19]; in fact, $D$ is both the diattenuation of $\mathbf{M}$ and the polarizance of the *reverse Mueller matrix* $\mathbf{M}^r \equiv \mathrm{diag}(1,1,-1,1)\,\mathbf{M}^T\,\mathrm{diag}(1,1,-1,1)$ [20,21] ($\mathbf{M}^T$ being the transposed matrix of $\mathbf{M}$) corresponding to the same interaction as $\mathbf{M}$ but interchanging the input and output directions. The mean intensity coefficient of $\mathbf{M}$ is given by $m_{00}$. For some purposes, it is useful to use the following normalized version of $\mathbf{M}$

$$\hat{\mathbf{M}} \equiv \mathbf{M}/m_{00} = \begin{pmatrix} 1 & \mathbf{D}^T \\ \mathbf{P} & \mathbf{m} \end{pmatrix}. \tag{2}$$

## 2. Components of purity

As a necessary step before further analyses, let us consider the notion of polarimetric purity. A given Mueller matrix **M** inherits, in some way, the statistical nature of the medium to which **M** is associated under given interaction conditions. Recall that a medium has a different associated Mueller matrix depending on 1) the spectral profile of the probe light beam; 2) the kind of interaction considered: refraction, reflection, scattering…; 3) the relative orientation of the medium with respect to the input beam; 4) the observation angle, etc. A medium that does not depolarize any totally polarized input beam is polarimetrically indistinguishable from a deterministic medium with well-defined Jones matrix **T** [and hence with well-defined *pure Mueller matrix* $\mathbf{M}_J(\mathbf{T})$ ] [5,22]. This kind of media is called *pure* or *nondepolarizing*. The closer is the polarimetric behavior to that of a nondepolarizing medium, the higher is its polarimetric purity.

A global measure of the *degree of polarimetric purity* of a medium is given by the *depolarization index* of **M** [16] defined as

$$P_\Delta = \sqrt{\left(D^2 + P^2 + 3P_S^2\right)/3}\,, \tag{3}$$

in terms of the *components of purity* (CP), namely the *polarizance P*, the *diattenuation D* and the *degree of spherical purity* $P_S \equiv \|\mathbf{m}\|_2/\sqrt{3}$ [2] ($\|\mathbf{m}\|_2$ representing the Euclidean norm of the $3\times3$ submatrix **m**). Conversely, an overall measure of the depolarizing power of a medium is given by the *depolarizance*

$$D_\Delta \equiv \sqrt{1-P_\Delta^2}\,. \tag{4}$$

Note that depolarizance was defined previously as $D_\Delta = 1 - P_\Delta$, but for reasons that are explained in Ref. [6], we consider more appropriate the indicated form.

Pure Mueller matrices are characterized by $P_\Delta = 1$ $\left(D_\Delta = 0\right)$, while Mueller matrices satisfying $P_\Delta < 1$ $\left(D_\Delta > 0\right)$ are called *nonpure* or *depolarizing* Mueller matrices. A medium satisfying $P_\Delta = 0$ $\left(D_\Delta = 1\right)$ converts any input polarization state into a fully depolarized output one. Despite the fact that $P_\Delta$ is an objective overall measure of the polarimetric purity (lack of randomness of the polarimetric properties of the interaction represented by **M**) it does not provide enough information for a complete parameterization of the polarimetric purity of **M**.

While *P* and *D* measure the relative portions of purity due to polarizance and diattenuation properties respectively, $P_S$ is a measure of the portion of purity that is not due to polarizance or diattenuation [2]. That is, the closer is $\hat{\mathbf{M}}$ to the Mueller matrix of a pure retarder (i.e., to an orthogonal Mueller matrix) the higher is the value of $P_S$. In fact, $P_S = 1$ if and only if $\hat{\mathbf{M}}$ is an orthogonal matrix. The value of $P_S$ is restricted to $0 \le P_S \le 1$, where the lower limit $P_S = 0$ is reached when the submatrix **m** is just the zero matrix and therefore the corresponding Mueller matrix has the form of an *absolute partial polarizer-analyzer*

$$\mathbf{M} = m_{00}\begin{pmatrix} 1 & \mathbf{D}^T \\ \mathbf{P} & \mathbf{0} \end{pmatrix}, \tag{5}$$

which is necessarily depolarizing since the minimum value of $P_S$ compatible with total purity of **M** is $P_S = 1/\sqrt{3}$ [2]. A detailed study of the achievable values for *P*, *D* and $P_S$ can be found in Ref [2].

Because of the common nature of *P* and *D*, and regardless the fact that they have respective specific and well-defined physical meanings, for some purposes it is useful to group them into the *degree of polarizance* [2]

$$P_P \equiv \sqrt{P^2 + D^2}/\sqrt{2}\,, \tag{6}$$

which is an overall measure of the polarizing power of the system represented by the Mueller matrix **M** (both forward and reverse incidence directions being considered). The value of $P_P$ is restricted to $0 \le P_P \le 1$, so that $P_P = 1$ corresponds to a total polarizer (the output states of both **M** and $\mathbf{M}^r$ are fully polarized regardless the degree of polarization of the input states). It should also be noted that pure diattenuators satisfy the condition $P_S^2 = 1 - 2P_P^2/3$ [2], so that a certain amount of spherical purity is consubstantial to this kind of systems. The value $P_P = 0$ is reached when the corresponding Mueller matrix **M** has zero polarizance and zero diattenuation.

Eq. (3) shows that the value of $P_\Delta$ is composed of the three complementary contributions of the corresponding *components of purity P*, *D* and $P_S$. Let us call *sources of purity* the quantities $P_P$ and $P_S$, which represent complementary contributions to the overall polarimetric purity given by $P_\Delta$

$$P_\Delta = \sqrt{\left(2P_P^2 + 3P_S^2\right)/3}\,. \tag{7}$$

### 3. Indices of polarimetric purity

A complementary approach for the study of the polarimetric purity of **M** is derived from the eigenvalue structure of the covariance matrix **H** defined as follows in terms of the elements $m_{kl}$ of **M**

$$\mathbf{H} = \frac{1}{4}\sum_{k,l=0}^{3} m_{kl}\left(\boldsymbol{\sigma}_k \otimes \boldsymbol{\sigma}_l\right), \tag{8}$$

where ⊗ stands for the Kronecker product and $\boldsymbol{\sigma}_k$ are the Pauli matrices

$$\boldsymbol{\sigma}_0 = \begin{pmatrix} 1 & 0 \\ 0 & 1 \end{pmatrix},\quad \boldsymbol{\sigma}_1 = \begin{pmatrix} 1 & 0 \\ 0 & -1 \end{pmatrix},\quad \boldsymbol{\sigma}_2 = \begin{pmatrix} 0 & 1 \\ 1 & 0 \end{pmatrix},\quad \boldsymbol{\sigma}_3 = \begin{pmatrix} 0 & -i \\ i & 0 \end{pmatrix}, \tag{9}$$

Let us now recall the notion of *parallel decomposition*, which consist of representing a Mueller matrix as a convex sum of Mueller matrices (i.e., a linear combination with positive coefficients that sum to one) [8]. The physical meaning of such decompositions is that the input electromagnetic beam is shared among a set of pencils that interact with a number of spatially distributed components in the illuminated area, in such a manner that the emerging pencils recombine incoherently into a single output beam.

Among the infinite possible parallel decompositions of **M** into pure Mueller matrices [8], let us bring up the *spectral decomposition* [7]

$$\mathbf{M} = \sum_{i=1}^{4} \hat{\lambda}_i \left(m_{00}\hat{\mathbf{M}}_{Ji}\right), \tag{10}$$

where $\hat{\lambda}_i \equiv \lambda_i/m_{00}$ are the ordered $\left(\hat{\lambda}_0 \ge \hat{\lambda}_1 \ge \hat{\lambda}_2 \ge \hat{\lambda}_3\right)$ normalized eigenvalues of **H**, and $\mathbf{M}_{Ji} \equiv m_{00}\hat{\mathbf{M}}_{Ji}$ are the pure Mueller matrices associated with the pure covariance matrices $\mathbf{H}_{Ji} \equiv m_{00}\left(\mathbf{u}_i \otimes \mathbf{u}_i^\dagger\right)$ defined from the respective eigenvectors $\mathbf{u}_i$ of **H**. Recall that $\operatorname{tr}\mathbf{H} = m_{00}$ [7] and that, since **H** is Hermitian and positive-semidefinite, $\lambda_i$ are nonnegative. Thus, the normalized eigenvalues $\hat{\lambda}_i$ of **H** have an immediate physical meaning as the relative weights of the pure components in the spectral decomposition. Moreover, the integer parameter $\operatorname{rank}\mathbf{H}$ gives the minimum number of parallel pure components of **M** [8]. An alternative and particularly significant way to express **M** as a convex sum of Mueller matrices is that provided by the *trivial*, or *characteristic*, decomposition [7,19]

$$\mathbf{M}(\mathbf{H}) = P_1 m_{00}\hat{\mathbf{M}}_{J1} + \left(P_2 - P_1\right)m_{00}\hat{\mathbf{M}}_2 + \left(P_3 - P_2\right)m_{00}\hat{\mathbf{M}}_3 + \left(1 - P_3\right)m_{00}\hat{\mathbf{M}}_{\Delta 0}, \tag{11}$$

where the quantities $P_1$, $P_2$ and $P_3$ are the so-called *indices of polarimetric purity* (IPP) of **M**, defined as follows from the normalized eigenvalues of **H** [1]

$$P_1 \equiv \hat{\lambda}_0 - \hat{\lambda}_1, \quad P_2 \equiv \hat{\lambda}_0 + \hat{\lambda}_1 - 2\hat{\lambda}_2, \quad P_3 \equiv \hat{\lambda}_0 + \hat{\lambda}_1 + \hat{\lambda}_2 - 3\hat{\lambda}_3, \tag{12}$$

and

- $\mathbf{M}_{J1} \equiv m_{00}\hat{\mathbf{M}}_{J1}$ is the *characteristic pure component*, determined by the eigenvector $\mathbf{u}_1$ associated with the largest eigenvalue of **H**. $\mathbf{M}_{J1}$ coincides with the first component of the spectral decomposition and its relative weight with respect to the complete system in Eq. (11) is given by the first index of polarimetric purity, or *degree of polarization* $P_1$ of **M**;
- $\mathbf{M}_2 \equiv m_{00}\hat{\mathbf{M}}_2$ represents a *2D depolarizer*, constituted by an equiprobable mixture of the first two spectral components. The relative weight of $\mathbf{M}_2$ in the trivial decomposition is given by the difference $P_2 - P_1$ between the first two IPP;
- $\mathbf{M}_3 \equiv m_{00}\hat{\mathbf{M}}_3$ represents a *3D depolarizer*, constituted by an equiprobable mixture of the first three spectral components. The relative weight of $\mathbf{M}_3$ in the trivial decomposition is given by the difference $P_3 - P_2$, and
- $\mathbf{M}_{\Delta 0} \equiv m_{00}\,\mathrm{diag}(1,0,0,0)$ represents an *ideal depolarizer* (or *4D depolarizer*), whose relative weight in the trivial decomposition is given by the difference $1 - P_3$.

The parameterization of the statistical relative weights in Eq. (11) by means of the IPP has the remarkable peculiarity that they satisfy the following nested inequalities

$$0 \le P_1 \le P_2 \le P_3 \le 1 \tag{13}$$

Therefore, regarding the structure of polarimetric purity in terms of statistical randomness, the IPP provide complete information, beyond the overall degree of polarimetric purity $P_\Delta$, which in fact is given by [1]

$$P_\Delta^2 = \frac{1}{3}\left(2P_1^2 + \frac{2}{3}P_2^2 + \frac{1}{3}P_3^2\right) \tag{14}$$

Thus, while the components of purity provide information of the polarimetric purity in terms of physical quantities related to the nature or "quality" of the medium, the IPP give it in terms of the structure of polarimetric randomness ("quantity of randomness"), but are insensitive to the nature of the sources of purity ($P_P$ and $P_S$). Note that the knowledge of the IPP does not imply necessarily the knowledge of $P$, $D$, and $P_S$, but the set of five quantities ($P_1, P_2, P_3, P, D$) is sufficient to calculate $P_\Delta$ and $P_S$. In fact, both sets of quantities are independent up to their respective quadratic relations with $P_\Delta$, so that, from Eqs. (3) and (14)

$$D^2 + P^2 + 3P_S^2 = 2P_1^2 + \frac{2}{3}P_2^2 + \frac{1}{3}P_3^2 \tag{15}$$

More details on the physical interpretation of both sets of parameters, can be found in previous works [1,2,19].

## 4. Invariant-equivalent Mueller matrices

As we have seen, given a Mueller matrix **M**, the corresponding values for its components of purity as well as for its IPP can be calculated from **M**. Nevertheless, the said quantities can correspond to different Mueller matrices and it is worth to identify the families that share the same values for $P_1, P_2, P_3, P, D, P_S$. To do so, let us consider the family of Mueller matrices $\mathbf{M}_E$ obtained through the pre- and post- multiplication of **M** by the Mueller matrices of arbitrary retarders $\mathbf{M}_{R2}$ and $\mathbf{M}_{R1}$ (recall that the Mueller matrix $\mathbf{M}_R$ of a retarder is an orthogonal Mueller matrix that has zero polarizance and diattenuation and whose submatrix $\mathbf{m}_R$ satisfies $\mathbf{m}_R^T = \mathbf{m}_R^{-1}$, with $\det \mathbf{m}_R = +1$) [19]

$$\begin{aligned}\mathbf{M}_E &\equiv m_{00}\mathbf{M}_{R2}\hat{\mathbf{M}}\mathbf{M}_{R1} = m_{00}\begin{pmatrix} 1 & \mathbf{0}^T \\ \mathbf{0} & \mathbf{m}_{R2}\end{pmatrix}\begin{pmatrix} 1 & \mathbf{D}^T \\ \mathbf{P} & \mathbf{m}\end{pmatrix}\begin{pmatrix} 1 & \mathbf{0}^T \\ \mathbf{0} & \mathbf{m}_{R1}\end{pmatrix} \\ &= m_{00}\begin{pmatrix} 1 & \mathbf{D}^T\mathbf{m}_{R1} \\ \mathbf{m}_{R2}\mathbf{P} & \mathbf{m}_{R2}\mathbf{m}\,\mathbf{m}_{R1}\end{pmatrix}\end{aligned} \tag{16}$$

Since $\mathbf{m}_R$ is an orthogonal matrix, the above *dual retarder transformation* [23], preserves the values of *P*, *D* and $P_S$. Furthermore, from the characteristic decomposition in Eq. (11),

$$\begin{gathered}\mathbf{M}_E \equiv m_{00}\mathbf{M}_{R2}\hat{\mathbf{M}}\mathbf{M}_{R1} = P_1 m_{00}\hat{\mathbf{M}}_{EJ1} + (P_2 - P_1)m_{00}\hat{\mathbf{M}}_{E2} + (P_3 - P_2)m_{00}\hat{\mathbf{M}}_{E3} + (1 - P_3)m_{00}\hat{\mathbf{M}}_{\Delta 0} \\ \hat{\mathbf{M}}_{EJ1} \equiv \mathbf{M}_{R2}\hat{\mathbf{M}}_{J1}\mathbf{M}_{R1}, \quad \hat{\mathbf{M}}_{E2} \equiv \mathbf{M}_{R2}\hat{\mathbf{M}}_{2}\mathbf{M}_{R1}, \quad \hat{\mathbf{M}}_{E3} \equiv \mathbf{M}_{R2}\hat{\mathbf{M}}_{3}\mathbf{M}_{R1}\end{gathered} \tag{17}$$

and consequently **M** and $\mathbf{M}_E$ also share the same values $P_1, P_2, P_3$ for their respective IPP.

Note that transformations of the form $\mathbf{M}_{J2}\mathbf{M}\mathbf{M}_{J1}$, where $\mathbf{M}_{J1}$ and $\mathbf{M}_{J2}$ are pure Mueller matrices having (either of them) nonzero polarizance-diattenuation, modify the nature of the components of the spectral and characteristic decompositions, so that the IPP are not preserved. Consequently, it is remarkable that the only transformations of **M** that preserve the IPP are the dual retarder transformations (as is the case with *P*, *D* and $P_S$).

Within the set of matrices $\mathbf{M}_E$ (which are called to be *invariant-equivalent* to **M**), let us bring up the so-called *arrow form* $\mathbf{M}_A$ of **M** **[24]**, whose definition is attained from the singular value decomposition of the submatrix **m** of **M**

$$\begin{gathered}\mathbf{m} = \mathbf{m}_{RO}\,\mathbf{m}_A\,\mathbf{m}_{RI}, \\ \mathbf{m}_{Ri}^{-1} = \mathbf{m}_{Ri}^T, \quad \det\mathbf{m}_{Ri} = +1 \quad (i = I, O), \\ \mathbf{m}_A \equiv \operatorname{diag}(a_1, a_2, \varepsilon a_3), \quad a_1 \geq a_2 \geq a_3 \geq 0, \quad \varepsilon \equiv (\det\mathbf{M})/|\det\mathbf{M}|,\end{gathered} \tag{18}$$

where the nonnegative parameters $(a_1, a_2, a_3)$ are the singular values of **m**, so that, by denoting the *entrance* and *exit* retarders as

$$\mathbf{M}_{Ri} = \begin{pmatrix} 1 & \mathbf{0}^T \\ \mathbf{0} & \mathbf{m}_{Ri}\end{pmatrix} \quad (i = I, O) \tag{19}$$

the *arrow matrix* $\mathbf{M}_A(\mathbf{M})$ is defined as [24]

$$\begin{gathered}\mathbf{M}_A(\mathbf{M}) \equiv \mathbf{M}_{RO}^T\,\mathbf{M}\,\mathbf{M}_{RI}^T = m_{00}\begin{pmatrix} 1 & \mathbf{D}_A^T \\ \mathbf{P}_A & \mathbf{m}_A\end{pmatrix} \\ \mathbf{D}_A = \mathbf{m}_{RI}\,\mathbf{D}, \quad \mathbf{P}_A = \mathbf{m}_{RO}^T\,\mathbf{P}\end{gathered} \tag{20}$$

Thus, the *arrow decomposition* of **M** is formulated as

$$\mathbf{M} = \mathbf{M}_{RO}\,\mathbf{M}_A\,\mathbf{M}_{RI} \tag{21}$$

Note that, to avoid ambiguity in the definition of $\mathbf{M}_A(\mathbf{M})$, the retarders $\mathbf{M}_{RI}$ and $\mathbf{M}_{RO}$ have been chosen so as to satisfy $a_1 \geq a_2 \geq a_3$ (and, hence, $1 \geq a_1 \geq a_2 \geq a_3 \geq 0$).

Since the six off-diagonal elements of $\mathbf{m}_A$ are zero, the main property of $\mathbf{M}_A(\mathbf{M})$ is that, besides it is invariant-equivalent to **M**, it contains, in a particularly simple and condensed manner, all the information concerning the ten physical invariants of **M** under dual retarder transformations. The set of ten parameters constituted by the mean transmittance $m_{00}$, together with the three vectors $\mathbf{D}_A$, $\mathbf{P}_A$ and the *sphericity vector* $\mathbf{P}_S \equiv (a_1, a_2, \varepsilon a_3)^T / \sqrt{3}$, contains complete information on the said invariant quantities [23] and, in particular, provide the CP, namely $|\mathbf{D}_A| = D$, $|\mathbf{P}_A| = P$ and $|\mathbf{P}_S| = P_S$.

In summary, regarding any analysis based on the structure and sources of polarimetric purity of **M** (fully determined by its IPP together with its CP), the arrow form $\mathbf{M}_A(\mathbf{M})$ is a particularly appropriate representative of **M** because it has, at least, six zero elements.

## 5. Purity figure

Once the two approaches for the representation of the purity structure of a Mueller matrix as well as the notion of invariant-equivalence between Mueller matrices have been analyzed, we are ready to address the study of the feasible regions of the space $P_P, P_S$ in terms of the IPP (Fig. 1).

The *purity figure*, determined by the feasible region in the graphic representation of $P_P$ versus $P_S$, is shown in Fig. 1 [2]. The space of the components of purity is located in the upper-right quadrant of the axes $P_S$ and $P_P$ and is limited on the right side by the elliptic branch $1 = 2P_P^2/3 + P_S^2$ and on the top side by the hyperbolic branch $P_P^2 = (1 + 3P_S^2)/2$ [2]. Let us first consider some general aspects of the purity figure.

- Axis $P_P = 0$ (segment OA): any parallel composition of retarders corresponds to this edge; also certain parallel combinations including diattenuators whose diattenuation-polarizance vectors are opposite, as for instance any arbitrary set of retarders combined with diattenuators such that the components can be grouped in the following way

$$\mathbf{M} = \frac{1}{2} m_{00} \begin{pmatrix} 1 & \mathbf{D}^T \\ \mathbf{P} & \mathbf{m}_1 \end{pmatrix} + \frac{1}{2} m_{00} \begin{pmatrix} 1 & -\mathbf{D}^T \\ -\mathbf{P} & \mathbf{m}_2 \end{pmatrix} \tag{22}$$

  A simple example of a Mueller matrix with $P_P = 0$ but containing diattenuators as parallel components is the following composition of two polarizers

$$\mathbf{M} = \frac{1}{2}\left[\frac{1}{2}\begin{pmatrix} 1 & -1 & 0 & 0 \\ 1 & -1 & 0 & 0 \\ 0 & 0 & 0 & 0 \\ 0 & 0 & 0 & 0 \end{pmatrix}\right] + \frac{1}{2}\left[\frac{1}{2}\begin{pmatrix} 1 & 1 & 0 & 0 \\ -1 & -1 & 0 & 0 \\ 0 & 0 & 0 & 0 \\ 0 & 0 & 0 & 0 \end{pmatrix}\right] = \frac{1}{2}\begin{pmatrix} 1 & 0 & 0 & 0 \\ 0 & -1 & 0 & 0 \\ 0 & 0 & 0 & 0 \\ 0 & 0 & 0 & 0 \end{pmatrix} \tag{23}$$

- Axis $P_S = 0$ (segment OD): any parallel composition of retarders and diattenuators that results in a Mueller matrix of the form of an absolute partial polarizer

$$\mathbf{M} = m_{00} \begin{pmatrix} 1 & \mathbf{D}^T \\ \mathbf{P} & \mathbf{0} \end{pmatrix} \tag{24}$$

  that is, the polarimetric effect of the medium is summarized into its polarizance and diattenuation vectors. A simple example of this kind of behavior can be synthesized by means of the following parallel composition of polarizers

$$\mathbf{M} = \frac{1}{2}\left[\frac{1}{2}\begin{pmatrix} 1 & 1 & 0 & 0 \\ 1 & 1 & 0 & 0 \\ 0 & 0 & 0 & 0 \\ 0 & 0 & 0 & 0 \end{pmatrix}\right] + \frac{1}{4}\left[\frac{1}{2}\begin{pmatrix} 1 & -1 & 0 & 0 \\ 1 & -1 & 0 & 0 \\ 0 & 0 & 0 & 0 \\ 0 & 0 & 0 & 0 \end{pmatrix}\right] + \frac{1}{4}\left[\frac{1}{2}\begin{pmatrix} 1 & 1 & 0 & 0 \\ -1 & -1 & 0 & 0 \\ 0 & 0 & 0 & 0 \\ 0 & 0 & 0 & 0 \end{pmatrix}\right] = \frac{1}{2}\begin{pmatrix} 1 & 1/2 & 0 & 0 \\ 1/2 & 0 & 0 & 0 \\ 0 & 0 & 0 & 0 \\ 0 & 0 & 0 & 0 \end{pmatrix} \tag{25}$$

- Segments OA and OD share the point O that corresponds to an ideal depolarizer $\mathbf{M}_{\Delta 0} = m_{00}\,\mathrm{diag}(1,0,0,0)$.

Different regions of the purity figure can be defined from the relative values of the IPP, namely: branch AC corresponds to $P_1 = 1$ (which, in turn, entails $P_2 = P_3 = P_\Delta = 1$); branch GD corresponds to $P_1 = 0$, $P_2 = 1$ (which, in turn, entails $P_3 = 1$); branch FE corresponds to $P_3 = 1$, $P_2 = 0$ (which entails $P_1 = 0$), and point O corresponds to $P_3 = 0$ (which entails $P_2 = P_1 = P_\Delta = 0$).

- *Pure media:* $P_1 = 1$ (hence, the three IPP are equal to 1). This maximum value of $P_1$ implies $\mathrm{rank}\,\mathbf{H} = 1$; $P_2 = P_3 = 1$; $P_\Delta = 1$, and $P_P^2 = 3\left(1 - P_S^2\right)/2$. The corresponding feasible region is the elliptical branch AC, $P_\Delta = 1$. From the point of view of the IPP, the system is composed of only one pure component, while the total purity of these states is shared among the sources of purity according to $2P_P^2 + 3P_S^2 = 3$. Any nondepolarizing Mueller matrix has an associated point on the line AC.

- *Two-dimensional media:* $1 = P_3 = P_2 > P_1 \geq 0$ (only two IPP are equal to 1). These values of the IPP correspond to $\mathrm{rank}\,\mathbf{H} = 2$ and also entail that the degree of polarimetric purity is limited by $1/\sqrt{3} \leq P_\Delta < 1$. Mueller matrices corresponding to this category of *two-dimensional media* are represented by points in the region ACDGA (curve AC excluded). From the point of view of the IPP, the system is equivalent to an incoherent composition of two pure media. The smaller is $P_1$, the closer is $P_\Delta$ to its lower limit $P_\Delta = 1/\sqrt{3}$, while the higher is the value of $P_1$, the closer is the value of $P_\Delta$ to $P_\Delta = 1$ (note that, as indicated above, the maximum value $P_\Delta = 1$ is exclusive of pure systems and thus it is not attained by depolarizing Mueller matrices, like those corresponding to two-dimensional media).

  From the point of view of the sources of purity, the overall purity $P_\Delta$ is shared among $P_S$ and $P_P$ according to $1/3 \leq 2P_P^2 + 3P_S^2 < 1$. As particular examples of media belonging to this type, we can mention parallel combinations of two retarders $\left(1/3 \leq 3P_S^2 < 1\right)$, or parallel combinations of two diattenuators $\left(1/3 \leq 2P_P^2 \leq 1\right)$. The closer are the optical axes of the components, the closer is $P_\Delta$ to $P_\Delta = 1$ (but, obviously, without reaching this limiting value). In general, any parallel composition of two pure media corresponds to this category.

- *Three-dimensional media:* $1 = P_3 > P_2 \geq P_1 \geq 0$ (only one IPP is equal to one). In this case, $\mathrm{rank}\,\mathbf{H} = 3$ and $1/3 \leq P_\Delta < 1$. The smaller are $P_2$ and $P_1$, the closer is $P_\Delta$ to its lower limit $P_\Delta = 1/3$, while the higher are $P_2$ and $P_1$, the closer is $P_\Delta$ to $P_\Delta = 1$ (but without reaching this limiting value). The corresponding depolarizing Mueller matrices are represented by points in the region ACDEFA (curve AC excluded). The values of the IPP determine that the system is equivalent to an incoherent composition of three pure media. Moreover, the overall purity $P_\Delta$ is shared among $P_S$ and $P_P$ according to $1/9 \leq 2P_P^2 + 3P_S^2 < 1$. As particular examples of media belonging to this type, we can mention parallel combinations of three retarders $\left(1/9 \leq 3P_S^2 < 1\right)$, or parallel combinations of three diattenuators $\left(1/9 \leq 2P_P^2 < 1\right)$. As is the case in any parallel composition of media of common nature (only retarders, only diattenuators…) the closer are the optical axes of the components, the closer to 1 is the overall purity.

- *Four-dimensional media:* $1 > P_3 \geq P_2 \geq P_1 \geq 0$ (none of the IPP reaches the maximum value 1). These values of the IPP imply that $\mathrm{rank}\,\mathbf{H} = 4$, while the degree of polarimetric purity is limited by $0 \leq P_\Delta < 1$. Region ACDOA (curve AC excluded), determine the points representing the corresponding Mueller matrices. This range of values of the IPP implies that the system is equivalent to an incoherent composition of four pure media. The overall purity $P_\Delta$ is shared among $P_S$ and $P_P$ according to $0 \leq 2P_P^2 + 3P_S^2 < 1$. As particular examples of media belonging to this type, we can mention parallel combinations of four retarders $\left(0 \leq 3P_S^2 < 1\right)$, or parallel combinations of four diattenuators $\left(0 \leq 2P_P^2 < 1\right)$. In general, parallel combinations of four or more different components correspond to this category. Note also that parallel combinations of *n* different components with $n > 4$ are polarimetrically indistinguishable from certain parallel combinations of four components. In fact any macroscopic interaction is constituted by a myriad of molecular effects, each of them with a well-defined Mueller matrix, but the overall polarimetric behavior (assuming linearity) is determined by the corresponding overall Mueller matrix, whose associated covariance has up to four nonzero eigenvalues (i.e., up to four degrees of freedom with respect to the equivalent pure components). Point O corresponds to an equiprobable mixture of four parallel components $(P_3 = 0)$, which is the so-called ideal depolarizer $\mathbf{M}_{\Delta 0} \equiv m_{00}\,\mathrm{diag}(1,0,0,0)$.

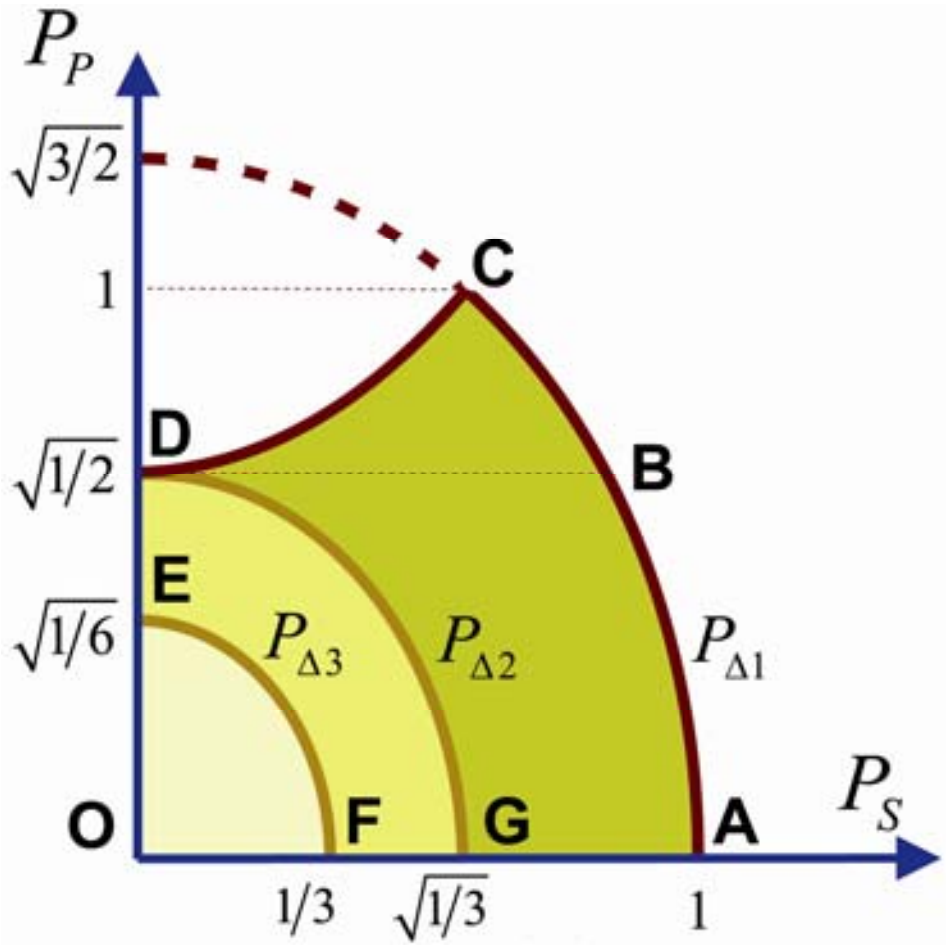

Fig. 1. $\mathrm{rank}\,\mathbf{H} \equiv r$ gives the minimum number of pure parallel components of $\mathbf{M}$. Different purity regions are determined by the integer value $r$ (which in turn is $r = 4 - \eta$, $\eta$ being the number of 1-valued IPP). Region I: curve AC ($r = 1$), which is exclusive for pure systems, whose total purity is shared among the sources of purity $P_P$ and $P_S$ (note that for pure systems $P = D = P_P$). Region II: ACDGA (branch AC excluded), determined by $1/\sqrt{3} \le P_\Delta < 1$, is feasible for systems with $r = 4,3,2$. Region III: GDEFG (branch GD excluded), determined by $1/3 \le P_\Delta < 1/\sqrt{3}$, is feasible for systems with $r = 4,3$. Region IV: FEOF (branch FE excluded), determined by $0 \le P_\Delta < 1/3$, is exclusive for systems with $r = 4$.

## 6. Purity figure for the canonical forms of depolarizing Mueller matrices

Any Mueller matrix $\mathbf{M}$ has an associated N-matrix defined as $\mathbf{N} \equiv \mathbf{G}\,\mathbf{M}^T\mathbf{G}\,\mathbf{M}$, where $\mathbf{G} \equiv (1,-1,-1,-1)$ is the Minkowski metric, and $\mathbf{M}$ is called type-I or type-II depending on whether $\mathbf{N}$ is diagonalizable or not. Furthermore, $\mathbf{M}$ can always be expressed as the following serial *symmetric decomposition* [25]

$$\mathbf{M} = \mathbf{M}_{J2}\,\mathbf{M}_\Delta\,\mathbf{M}_{J1}, \tag{26}$$

where $\mathbf{M}_{J1}$ and $\mathbf{M}_{J2}$ are pure Mueller matrices, while $\mathbf{M}_\Delta$ represents a *canonical depolarizer* that, in the case of $\mathbf{M}$ being type-I, has the form [15,25]

$$\begin{gathered}\mathbf{M}_{\Delta d} \equiv \mathrm{diag}(d_0, d_1, d_2, \varepsilon d_3);\\ d_0 \ge d_i \ge 0 \quad (i = 1,2,3), \qquad \varepsilon \equiv (\det\mathbf{M})/|\det\mathbf{M}|,\\ d_0 + d_1 \ge d_2 + \varepsilon d_3, \qquad d_0 - d_1 \ge d_2 - \varepsilon d_3,\end{gathered} \tag{27}$$

and when $\mathbf{M}$ is type-II, has the alternative canonical form [3]

$$\mathbf{M}_{\Delta nd} \equiv \begin{pmatrix} 2a_0 & -a_0 & 0 & 0 \\ a_0 & 0 & 0 & 0 \\ 0 & 0 & a_2 & 0 \\ 0 & 0 & 0 & a_2 \end{pmatrix}; \quad (0 \le a_2 \le a_0) \tag{28}$$

The parameters $(d_0, d_1, d_2, d_3)$ and $(a_0, a_0, a_2, a_2)$ are the square roots of the (nonnegative) eigenvalues $(\rho_0, \rho_1, \rho_2, \rho_3)$ of $\mathbf{N}$. When appropriate we will use $\mathbf{M}_\Delta$ to refer to the Mueller matrix of the canonical depolarizer without specifying if it is type-I or type-II.

Prior to analyze the respective purity figures for $\mathbf{M}_{\Delta d}$ and $\mathbf{M}_{\Delta nd}$, let us note that they are not invariant-equivalent to $\mathbf{M}$, and their IPP are, in general, different from those of the composed Mueller matrix $\mathbf{M}$ [26]. Nevertheless, it is remarkable that, since the ranks of the covariance matrices associated with $\mathbf{M}$ and $\mathbf{M}_\Delta$ are equal $\left(\mathrm{rank}\left[\mathbf{H}(\mathbf{M}_\Delta)\right] = \mathrm{rank}\left[\mathbf{H}(\mathbf{M})\right]\right)$, the number $\eta$ of IPP that are equal to one (where $\eta = 4 - \mathrm{rank}\left[\mathbf{H}(\mathbf{M})\right]$) also coincides for both matrices (regardless of whether $\mathbf{M}$ is type-I or type-II). Obviously, the components of purity of $\mathbf{M}_\Delta$ are in general different from those of $\mathbf{M}$ and in the next subsections we will consider the purity figures for $\mathbf{M}_{\Delta d}$ and $\mathbf{M}_{\Delta nd}$ taken as individual matrices and not for Mueller matrices containing them as serial components.

### *6.1 Purity figure for the type-I canonical depolarizer*

The IPP of $\mathbf{M}_{\Delta d}$ can be expressed as follows in terms of its diagonal elements $d_i$ [26]

$$P_1 = \frac{d_2 + \varepsilon d_3}{2d_0}, \quad P_2 = \frac{2d_1 - d_2 + \varepsilon d_3}{2d_0}, \quad P_3 = \frac{d_1 + d_2 - \varepsilon d_3}{d_0} \tag{29}$$

so that the general form $\mathbf{M}_{\Delta d}$ in terms of the IPP is

$$\mathbf{M}_{\Delta d} = d_0\, \mathrm{diag}\left(1, \frac{2P_2 + P_3}{3}, P_1 + \frac{P_3 - P_2}{3}, \varepsilon P_1 - \varepsilon \frac{P_3 - P_2}{3}\right) \tag{30}$$

Due to the simple diagonal structure of $\mathbf{M}_{\Delta d}$, its components of purity adopt the following values

$$\begin{gathered} D(\mathbf{M}_{\Delta d}) = P(\mathbf{M}_{\Delta d}) = P_P(\mathbf{M}_{\Delta d}) = 0 \\ 0 \le P_S(\mathbf{M}_{\Delta d}) = P_\Delta(\mathbf{M}_{\Delta d}) = \sqrt{\frac{d_1^2 + d_2^2 + d_3^2}{3d_0^2}} \le 1 \end{gathered} \tag{31}$$

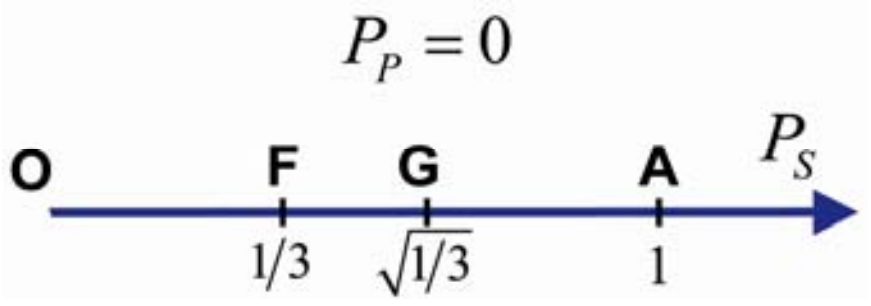

Fig. 2. The purity figure for $\mathbf{M}_{\Delta d}$ is determined by the axis $P_P = 0$. Region I: Point A ($\mathrm{rank}\mathbf{H} = 1$), is exclusive for pure systems with $P_S = 1$ (pure retarders). Region II: segment GA (point A excluded), determined by $1/\sqrt{3} \le P_\Delta = P_S < 1$, is feasible for systems with $r = 4,3,2$. Region III: segment FG (point G excluded), determined by $1/3 \le P_\Delta = P_S < 1/\sqrt{3}$, is feasible for systems with $r = 4,3$. Region IV: segment OF (point F excluded), determined by $0 \le P_\Delta = P_S < 1/3$, is exclusive for systems with $r = 4$.

Therefore, the feasible points representing $\mathbf{M}_{\Delta d}$ in the purity figure are arranged along the $P_S$ axis (Fig. 2) and according with the structure of its IPP, we can distinguish the following cases

- *Neutral filter:* $P_1 = 1$. $\mathbf{M}_{\Delta d}$ has the simple form $\mathbf{M}_{\Delta d} = d_0 \mathbf{I}$, (**I** being the identity matrix) and it is represented by point A in the purity figure, corresponding to $P_S = 1$, $P_P = 0$.

- *Two-dimensional type-I canonical depolarizer:* $P_2 = 1$, $P_1 < 1$. $\mathbf{M}_{\Delta d}$ adopts the form $\mathbf{M}_{\Delta d} = d_0 \,\mathrm{diag}\left(1,1,P_1,P_1\right)$, so that the purity parameters are given by

$$P_P = 0, \quad 1\big/\sqrt{3} \le P_S = P_\Delta = \sqrt{\frac{d_0^2 + 2d_2^2}{3d_0^2}}, \qquad P_1 = d_2 / d_0 < 1, \quad P_2 = P_3 = 1. \tag{32}$$

  The feasible region is determined by the segment GA (point A excluded), and point G corresponds to the particular case $d_2 = 0$. The explicit expressions for the arbitrary parallel decompositions of $\mathbf{M}_{\Delta d}$ as convex linear combinations of pure Mueller matrices can be found in Ref. [27].

- *Three-dimensional type-I canonical depolarizer:* $P_3 = 1$, $P_2 < 1$. $\mathbf{M}_{\Delta d}$ adopts the form

$$\mathbf{M}_{\Delta d} = d_0 \,\mathrm{diag}\left(1, \frac{1+2P_2}{3}, P_1 + \frac{1-P_2}{3}, \varepsilon P_1 - \varepsilon \frac{1-P_2}{3}\right) \tag{33}$$

  so that the purity parameters are now

$$P_P = 0, \quad 1/3 < P_S = P_\Delta = \sqrt{\frac{d_1^2 + d_2^2 + \left(d_1 + d_2 - d_0\right)^2}{3d_0^2}} < 1, \qquad 0 \le P_1 = \frac{d_1 + 2d_2 - d_0}{2d_0} \le P_2 = \frac{3d_1 - d_0}{2d_0} < P_3 = 1. \tag{34}$$

  In this case, $\mathbf{M}_{\Delta d}$ is represented by points located in the segment FA (point A excluded). Point F corresponds to $P_1 = P_2 = 0$ $\left(d_1 = d_2 = -d_3 = d_0/3\right)$ and $P_S = P_\Delta = 1/3$. Point G corresponds to $P_1 = P_2 = 1/2$ $\left(d_1 = d_2 = 2d_3 = 2d_0/3\right)$ and $P_S = P_\Delta = 1\big/\sqrt{3}$.

  The explicit expressions for the arbitrary parallel decompositions of $\mathbf{M}_{\Delta d}$ as convex linear combinations of pure Mueller matrices can be found in Ref. [27].

- *Four-dimensional type-I canonical depolarizer:* $P_3 < 1$. $\mathbf{M}_{\Delta d}$ adopts the form (30) and the purity parameters take the forms

$$P_P = 0, \quad 0 \le P_S = P_\Delta = \sqrt{\left(d_1^2 + d_2^2 + d_3^2\right)\big/3d_0^2} < 1, \qquad 0 \le P_1 = \frac{d_1 + d_3}{2d_0} \le P_2 = \frac{2d_1 - d_2 + d_3}{2d_0} \le P_3 = \frac{d_1 + d_2 - d_3}{d_0} < 1. \tag{35}$$

  The four-dimensional $\mathbf{M}_{\Delta d}$ is represented by points located in the segment OA (point A excluded). Point O corresponds to an ideal depolarizer, with $P_1 = P_2 = P_3 = 0$ $\left(d_1 = d_2 = d_3 = 0\right)$ and $P_S = P_\Delta = 0$. Point F corresponds to $P_1 = P_2 = P_3 = 1\big/\sqrt{3}$ $\left(d_1 = d_2 = d_3 = d_0\big/\sqrt{3}\right)$ and $P_S = P_\Delta = 1/3$. Point G corresponds to $P_1 = P_2 = P_3 = 1/3$ $\left(d_1 = d_2 = d_3 = d_0/3\right)$ and $P_S = P_\Delta = 1\big/\sqrt{3}$.

### *6.2 Purity figure for the type-II canonical depolarizer*

Due to the particular structure of $\mathbf{M}_{\Delta nd}$ the quantities $D, P, P_P, P_S$ and $P_\Delta$ take the following values

$$D\left(\mathbf{M}_{\Delta nd}\right)=P\left(\mathbf{M}_{\Delta nd}\right)=P_P\left(\mathbf{M}_{\Delta nd}\right)=\frac{1}{2},$$

$$0\le P_S\left(\mathbf{M}_{\Delta nd}\right)=\frac{a_2}{\sqrt{6}a_0}\le\frac{1}{\sqrt{6}}, \tag{36}$$

$$\frac{1}{\sqrt{6}}\le P_\Delta\left(\mathbf{M}_{\Delta nd}\right)=\sqrt{\frac{a_0^2+a_2^2}{6a_0^2}}\le\frac{1}{\sqrt{3}},$$

and the IPP of $\mathbf{M}_{\Delta nd}$ can be expressed as follows in terms of $a_0$ and $a_2$

$$0\le P_1=\frac{1}{4a_0}\left(a_0-a_2\right)\le\frac{1}{4},$$

$$\frac{1}{4}\le P_2=\frac{1}{4a_0}\left(a_0+3a_2\right)\le 1, \tag{37}$$

$$P_3=1.$$

so that the general form of $\mathbf{M}_{\Delta nd}$ in terms of its IPP is

$$\mathbf{M}_{\Delta nd}=a_0\begin{pmatrix}2 & -1 & 0 & 0\\ 1 & 0 & 0 & 0\\ 0 & 0 & P_2-P_1 & 0\\ 0 & 0 & 0 & P_2-P_1\end{pmatrix}, \tag{38}$$

with the restrictions for $P_1$ and $P_2$ indicated in Eq. (37). Note that, since $P_3=1$, $\mathbf{M}_{\Delta nd}$ cannot be four-dimensional (that is $\operatorname{rank}\left[\mathbf{H}\left(\mathbf{M}_{\Delta nd}\right)\right]<4$), and since $P_1\le 1/4$ $\mathbf{M}_{\Delta nd}$ cannot be pure [3].

The feasible region for $\mathbf{M}_{\Delta nd}$ in the purity figure is given by segment HI (Fig. 3) and, according with the structure of its IPP, we can distinguish the two following cases

- *Two-dimensional type-II canonical depolarizer:* $P_2=1, P_1=0$. $\mathbf{M}_{\Delta nd}$ adopts the form

$$\mathbf{M}_{\Delta nd}=a_0\begin{pmatrix}2 & -1 & 0 & 0\\ 1 & 0 & 0 & 0\\ 0 & 0 & 1 & 0\\ 0 & 0 & 0 & 1\end{pmatrix}, \tag{39}$$

  so that the purity parameters are given by $D=P=P_P=1/2$, $P_S=1/\sqrt{6}$, $P_\Delta=1/\sqrt{3}$, $P_1=0$, $P_2=P_3=1$. In this case $\mathbf{M}_{\Delta nd}$ is represented by the single point I in the purity figure (Fig. 3).

- *Three-dimensional type-II canonical depolarizer:* $P_2<1$. $\mathbf{M}_{\Delta nd}$ adopts the generic form (38), with the restriction $P_2<1$ added to conditions (37), and $\mathbf{M}_{\Delta nd}$ is represented by points in the segment HI (point I excluded) of the purity figure (Fig. 3). Point H corresponds to the particular case $P_2=P_1$ $\left(a_2=0\right)$.

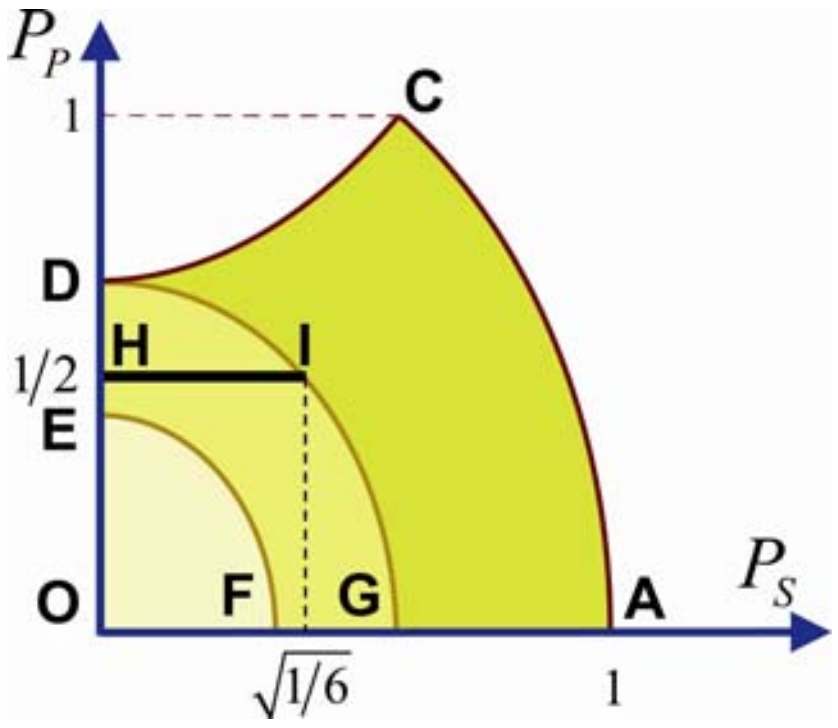

Fig. 3. Segment HI is the feasible region for canonical type-II Mueller matrices.

## 7. Parameterization of a general Mueller matrix in terms of depolarization, retardance, diattenuation and polarizance parameters

From the arrow decomposition $\mathbf{M} = \mathbf{M}_{RO}\,\mathbf{M}_A\,\mathbf{M}_{RI}$, [Eq. (21)] the sixteen independent parameters involved in **M** can be parameterized as follows:

- The mean intensity coefficient $m_{00}$ of **M** (which coincides with that of $\mathbf{M}_A$).
- Three parameters of the *entrance retarder* $\mathbf{M}_{RI}$, namely the azimuth $\varphi_I$ and ellipticity angle $\chi_I$ of the fast eigenstate of $\mathbf{M}_{RI}$ (which in general is an elliptic retarder), together with the retardance $\Delta_I = \arccos\left[\left(\operatorname{tr}\mathbf{M}_{RI} - 2\right)/2\right]$ introduced by $\mathbf{M}_{RI}$ between its fast and slow eigenstates. These parameters determine the *straight retardance vector* $\bar{\mathbf{R}}_I$ of $\mathbf{M}_{RI}$, (or *entrance retardance vector* of **M**) defined as

$$\bar{\mathbf{R}}_I = \left(\Delta_I/\pi\right)\left(\cos 2\varphi_I \cos 2\chi_I, \sin 2\varphi_I \cos 2\chi_I, \sin 2\chi_I\right)^T \tag{40}$$

- Three parameters of the *exit retarder* $\mathbf{M}_{RO}$, namely the azimuth $\varphi_O$ and ellipticity angle $\chi_O$ of the fast eigenstate of $\mathbf{M}_{RO}$ (which in general is an elliptic retarder), together with the retardance $\Delta_O = \arccos\left[\left(\operatorname{tr}\mathbf{M}_{RO} - 2\right)/2\right]$. These parameters determine the straight retardance vector $\bar{\mathbf{R}}_O$ of $\mathbf{M}_{RO}$, (or *exit retardance vector* of **M**) defined as

$$\bar{\mathbf{R}}_O = \left(\Delta_O/\pi\right)\left(\cos 2\varphi_O \cos 2\chi_O, \sin 2\varphi_O \cos 2\chi_O, \sin 2\chi_O\right)^T \tag{41}$$

- Three parameters of the *diattenuation vector* **D** of **M**

$$\mathbf{D} = \left(1/m_{00}\right)\left(m_{01}, m_{02}, m_{03}\right)^T = \mathbf{m}_{RI}^T\,\mathbf{D}_A \tag{42}$$

- Three parameters of the *polarizance vector* **P** of **M**

$$\mathbf{P} = \left(1/m_{00}\right)\left(m_{10}, m_{20}, m_{30}\right)^T = \mathbf{m}_{RO}\,\mathbf{P}_A \tag{43}$$

- The three indices of polarimetric purity $P_1, P_2, P_3$ of **M** (which coincide with those of $\mathbf{M}_A$).

The above analysis stresses the fact that the structure of polarimetric randomness of **M** is fully determined by the IPP. When $P_1 = 1$ (i.e. $P_\Delta = 1$, and consequently **M** is a pure Mueller matrix $\mathbf{M}_J$), $\mathbf{M}_A$ takes the form of a normal [28,19] (or *homogeneous* [29]) linear diattenuator oriented at 0º [19]

$$\mathbf{M}_A\left(P_1=1\right)=\mathbf{M}_{DL0}\equiv m_{00}\begin{pmatrix}1 & \cos\kappa & 0 & 0\\ \cos\kappa & 1 & 0 & 0\\ 0 & 0 & \sin\kappa & 0\\ 0 & 0 & 0 & \sin\kappa\end{pmatrix}, \qquad (44)$$
$$\left(0\le\kappa\le\pi/2\right),$$

where $D\equiv\cos\kappa$ is the diattenuation-polarizance of $\mathbf{M}_J$ and $\sqrt{1-D^2}\equiv\sin\kappa$ can be properly called the *counterpolarizance* of $\mathbf{M}_J$. Then, for this case of a pure system, the arrow decomposition becomes the *general serial decomposition* of $\mathbf{M}_J$ [5]

$$\mathbf{M}_J\equiv m_{00}\begin{pmatrix}1 & \mathbf{D}^T\\ \mathbf{P} & \mathbf{m}\end{pmatrix}=\mathbf{M}_{RO}\,\mathbf{M}_{DL0}\,\mathbf{M}_{RI}$$
$$=\begin{pmatrix}1 & \mathbf{0}^T\\ \mathbf{0} & \mathbf{m}_{RO}\end{pmatrix}\left\{m_{00}\begin{pmatrix}1 & \mathbf{D}_0^T\\ \mathbf{D}_0 & \mathbf{m}_{D_0}\end{pmatrix}\right\}\begin{pmatrix}1 & \mathbf{0}^T\\ \mathbf{0} & \mathbf{m}_{RI}\end{pmatrix}=m_{00}\begin{pmatrix}1 & \left(\mathbf{m}_{RI}^T\mathbf{D}_0\right)^T\\ \mathbf{m}_{RO}\mathbf{D}_0 & \mathbf{m}_{RO}\mathbf{m}_{D_0}\mathbf{m}_{RI}\end{pmatrix} \qquad (45)$$
$$\left[\mathbf{D}_0\equiv\left(\cos\kappa,0,0\right)^T,\ \mathbf{m}_{D_0}\equiv\operatorname{diag}\left(1,\sin\kappa,\sin\kappa\right)\right]$$

where,

$$\mathbf{D}=\mathbf{m}_{RI}^T\mathbf{D}_0 \qquad \mathbf{P}=\mathbf{m}_{RO}\mathbf{D}_0 \qquad \mathbf{m}=\mathbf{m}_{RO}\mathbf{m}_{D_0}\mathbf{m}_{RI} \qquad (46)$$

showing that $\mathbf{m}_{RI}$ and $\mathbf{m}_{RO}^T$ rotate respectively the diattenuation vector $\mathbf{D}$ and the polarizance vector $\mathbf{P}$ until they coincide with $\mathbf{D}_0\equiv\left(D,0,0\right)^T$.

Since $\mathbf{M}_J$ depends on up to seven independent parameters, decomposition (45) can always be expressed as

$$\mathbf{M}_J=\left[\mathbf{M}_{RLO}\left(\varphi_{LO},\Delta_{LO}\right)\mathbf{M}_{RL0}\left(\Delta/2\right)\right]\mathbf{M}_{DL0}\left(m_{00},D\right)\left[\mathbf{M}_{RL0}\left(\Delta/2\right)\mathbf{M}_{RLI}\left(\varphi_{LI},\Delta_{LI}\right)\right] \qquad (47)$$

where

$$\mathbf{M}_{Ri}\left(\varphi_i,\Delta_i\right)=\begin{pmatrix}1 & 0 & 0 & 0\\ 0 & \cos^2 2\varphi_i+\cos\Delta_i\,\sin^2 2\varphi_i & \left(1-\cos\Delta_i\right)\sin 2\varphi_i\,\cos 2\varphi_i & -\sin 2\varphi_i\,\sin\Delta_i\\ 0 & \left(1-\cos\Delta_i\right)\sin 2\varphi_i\,\cos 2\varphi_i & \sin^2 2\varphi_i+\cos\Delta_i\,\cos^2 2\varphi_i & \cos 2\varphi_i\,\sin\Delta_i\\ 0 & \sin 2\varphi_i\,\sin\Delta_i & -\cos 2\varphi_i\,\sin\Delta_i & \cos\Delta_i\end{pmatrix} \qquad (48)$$

with $\left(i=LI,LO\right)$, $\mathbf{M}_{RLI}\left(\varphi_{LI},\Delta_{LI}\right)$ and $\mathbf{M}_{RLO}\left(\varphi_{LO},\Delta_{LO}\right)$ being the Mueller matrices of the *entrance* and *exit* linear retarders with respective azimuths $\left(\varphi_{LI},\varphi_{LO}\right)$ and retardances $\left(\Delta_{LI},\Delta_{LO}\right)$, while

$$\mathbf{M}_{RL0}\left(\Delta/2\right)=\begin{pmatrix}1 & 0 & 0 & 0\\ 0 & 1 & 0 & 0\\ 0 & 0 & \cos\left(\Delta/2\right) & \sin\left(\Delta/2\right)\\ 0 & 0 & -\sin\left(\Delta/2\right) & \cos\left(\Delta/2\right)\end{pmatrix} \qquad (49)$$

is the Mueller matrix of a linear retarder oriented at 0º with retardance $\Delta/2$ Therefore, the retardance properties of the pure Mueller matrix $\mathbf{M}_J$ are fully determined by the pair of retarders

$$\begin{aligned}\mathbf{M}_{RI}&=\mathbf{M}_{RL0}\left(\Delta/2\right)\mathbf{M}_{RLI}\left(\varphi_{LI},\Delta_{LI}\right)\\ \mathbf{M}_{RO}&=\mathbf{M}_{RLO}\left(\varphi_{LO},\Delta_{LO}\right)\mathbf{M}_{RL0}\left(\Delta/2\right)\end{aligned} \qquad (50)$$

whose five characteristic parameters, together with the mean intensity coefficient $m_{00}$ and the diattenuation-polarizance *D*, constitute a complete set $\left(m_{00}, D, \varphi_{LI}, \varphi_{LO}, \Delta_{LI}, \Delta_{LO}, \Delta\right)$ of seven independent quantities characterizing the polarimetric properties of $\mathbf{M}_J$.

The above result highlights the fact that the structure of polarimetric randomness is fully determined by the three IPP, which characterize the depolarization properties and that when the system is nondepolarizing, the components of the arrow decomposition adopt the above indicated simple forms $\mathbf{M}_{RI} = \mathbf{M}_{RL0}\,\mathbf{M}_{RLI}$, $\mathbf{M}_{DL0}$ and $\mathbf{M}_{RO} = \mathbf{M}_{RLO}\,\mathbf{M}_{RL0}$, so that the number of 16 free parameters of a general Mueller matrix is reduced to 7 for a pure Mueller matrix.

Concerning polarizance *P* of **M**, let us recall that media with $P > 0$ increase the degree of polarization of certain partially polarized input states, but it should be stressed that such media also decrease the degree of polarization for some partially polarized input states [30]. Furthermore, since the diattenuation *D* of **M** is also the polarizance of $\mathbf{M}^r$, the essential physical nature of *D* is analogous to that of *P* and consequently, depolarization, polarizance and diattenuation have entangled natures. The amount of depolarization is fully determined by the IPP, while the analysis of the sources (or "quality") of depolarization requires to consider the three CP.

It is worth to observe that an alternative parameterization of a pure Mueller matrix is achieved through the polar decomposition [4]

$$\mathbf{M}_J = \mathbf{M}_R\mathbf{M}_D = \mathbf{M}_P\mathbf{M}_R \tag{51}$$

where the diattenuators $\mathbf{M}_D$ and $\mathbf{M}_P$ of the two forms are mutually related through $\mathbf{M}_D = \mathbf{M}_R^T\mathbf{M}_P\mathbf{M}_R$ and where $\mathbf{M}_R$ (which is the same for the two above indicated alternative forms of the polar decomposition) can apparently be considered as a retarder that summarizes the effective overall retardation properties of the pure Mueller matrix $\mathbf{M}_J$. Nevertheless, $\mathbf{M}_D$ and $\mathbf{M}_P$ are in general different, and each of them can be decomposed as

$$\begin{aligned}\mathbf{M}_D &= \mathbf{M}_{Rf}\,\mathbf{M}_{DL0}\,\mathbf{M}_{Rf}^T\\ \mathbf{M}_P &= \mathbf{M}_{Rr}\,\mathbf{M}_{DL0}\,\mathbf{M}_{Rr}^T\end{aligned} \tag{52}$$

so that both forward ($\mathbf{M}_D$) and reverse ($\mathbf{M}_P$) diattenuators involve consubstantial respective retarders $\mathbf{M}_{Rf}$ and $\mathbf{M}_{Rr}$, and therefore, unlike what it seems at first glance, the retardance parameters of $\mathbf{M}_R$ do not provide complete and decoupled information on the overall retardance properties of $\mathbf{M}_J$, which in fact are properly determined by $\left(\varphi_{LI}, \varphi_{LO}, \Delta_{LI}, \Delta_{LO}, \Delta\right)$.

## 8. Discussion and conclusions

Polarimetric purity refers to the closeness of the polarimetric behavior of a material sample to a deterministic one (hence nondepolarizing) and, therefore, this notion is intimately linked to the depolarization properties of the sample under consideration. The closer is the depolarizance $D_\Delta$ to its maximum achievable value $D_\Delta = 1$ (i.e. $P_\Delta = 0$), the closer is the behavior of the medium to an ideal depolarizer, which in turn is polarimetrically equivalent to an equiprobable parallel mixture of its four spectral components. The closer is $D_\Delta$ to its minimum achievable value $D_\Delta = 0$ (i.e. $P_\Delta = 1$), the closer is the behavior to a deterministic one, which constitutes, by itself, its only nonvanishing spectral component. In summary, polarimetric purity is related to two kinds of properties of the medium, namely *1)* the lack of randomness of its spectral statistical structure, and *2)* the lack of ability to depolarize the input totally polarized electromagnetic beams.

While an overall measure of the degree of polarimetric purity is given by the depolarization index $P_\Delta$, this invariant quantity, taken alone, does not provide complete information on the structure of polarimetric randomness (i.e. on the relative weights of the spectral components), nor on the polarizance, diattenuation and spherical purity of the medium. Nevertheless, the depolarization properties can be fully characterized through two complementary views, namely:

- The three IPP, which give complete information on the structure of polarimetric randomness of the sample in terms of the relative weights of the components of the characteristic decomposition of the Mueller matrix **M**, but which are insensitive to the polarizance, diattenuation and spherical properties of the sample. Note, in passing, that the integer quantity $\mathrm{rank}\mathbf{H}(\mathbf{M}) \equiv r$ (which in turn is given by $r = 4 - \eta$, $\eta$ being the number of 1-valued IPP) is equal to the minimum number of pure parallel components of the system.
- The three CP, which give complete information on the structure of polarimetric purity in terms of the contributions due to the degree of spherical purity$(P_S)$, the polarizance $(P)$ and the diattenuation $(D)$.

By combining the sets of IPP and CP we obtain the following set of five non-dimensional independent quantities $(D, P, P_1, P_2, P_3)$ (recall that $P_S$ can be calculated from this set), all of them limited to values between 0 and 1, which fully determine, in quantity and quality, the integral depolarizing properties of **M**. The said quantities $(D, P, P_1, P_2, P_3)$, as well as other parameters obtainable from them, like $(r, P_S, P_P, P_\Delta, D_\Delta)$, are intrinsic of the system (for the given conditions of interaction with polarized light) in the sense that they are invariant under changes of generalized bases of representation of the input and output states of polarization of light [2]. In other words, these *intrinsic quantities* of **M** are invariant when **M** is pre- or post-multiplied by orthogonal Mueller matrices (retarders) [23].

Due to the common physical nature of polarizance *P* and diattenuation *D*, the quantity $P_P$ (degree of polarizance) is useful for representing the combined contributions of *P* and *D* versus the degree of spherical purity $P_S$. This representation defines a "purity figure" where the feasible regions for systems with different values of *r* can be identified in terms of the IPP. The purity figures for the canonical type-I and type-II depolarizers (which play a key role in Mueller algebra) have been identified as the two respective segments shown in Fig. 2 and Fig. 3 respectively.

Furthermore, the arrow decomposition $\mathbf{M} = \mathbf{M}_{RO}\,\mathbf{M}_A\,\mathbf{M}_{RI}$, [Eq. (21)] provides a meaningful parameterization of the sixteen independent real quantities involved in **M**, namely the mean intensity coefficient $m_{00}$, the components of the straight retardance vectors $\bar{\mathbf{R}}_I$ and $\bar{\mathbf{R}}_O$, the components of the diattenuation and polarizance vectors **D** and **P**, and the three IPP $P_1, P_2, P_3$. For a nondepolarizing medium these quantities collapse into a set of seven independent parameters $(m_{00}, D, \varphi_{LI}, \varphi_{LO}, \Delta_{LI}, \Delta_{LO}, \Delta)$ instead of the sixteen independent parameters involved in a general depolarizing Mueller matrix.

*The author is grateful to the referee for his pertinent suggestions that helped improve the quality of the manuscript.*

## Funding

Ministerio de Economía y Competitividad (FIS2011-22496 and FIS2014-58303-P); Gobierno de Aragón (E99).